\documentstyle[prb,aps,tighten,eqsecnum,floats,epsf]{revtex}
\preprint{T97/072} 
\begin{document} 
\draft 
\title{EFFECTS OF RANDOMNESS ON THE ANTIFERROMAGNETIC SPIN-1 CHAIN}
\author{V. Brunel and Th. Jolic\oe ur\thanks{e-mail: vbrunel,
thierry@spht.saclay.cea.fr}} 
\address{Service de Physique
Th\'eorique, CEA Saclay,\\ F91191 Gif-sur-Yvette, France}
\date{January 19, 1998} 
\maketitle
\begin{abstract} 
We investigate the effect of weak randomness 
on the antiferromagnetic anisotropic spin-1
chain. We use Abelian
bosonization to construct the low-energy effective theory.
A renormalization group calculation up to second order
in the strength of the disorder is performed on this effective 
theory. 
We observe in this framework the destruction
of the antiferromagnetic ordered phase {\`a la} Imry-Ma.
We predict the effects of a random magnetic field along z 
axis, a random field in the XY plane as well as random exchange
with and without XY symmetry. 
Instabilities of massless phases appear in general 
by mechanisms different from the case of the 2-leg spin ladder.
\end{abstract}
\section{Introduction}

Quantum one-dimensional spin systems are characterized by a variety
of behavior much richer than their higher-dimensional counterpart.
For example, antiferromagnetic spin chains have different physical
properties according to whether the spins are integer or
half-integer\cite{FDM}. Let us consider the prototypical
Heisenberg spin chain~:
\begin{equation} 
{\mathcal H} = J \sum_i {\bf S_{i}}\cdot {\bf S_{i+1}},
 \label{Heis}
\end{equation} 
where i are the sites along the chain and $\bf S_{i}$ quantum spin
operators with ${\bf S_{i}}^{2}=S(S+1)$. 
The exchange coupling $J$ is positive, i.e. antiferromagnetic.
Then the S=1/2 chain is gapless and has
algebraic decay of the spin correlations 
$\langle {\bf S_{0}}\cdot {\bf S_{n}}\rangle \simeq (-)^{n}/n$
for $n\rightarrow\infty$. 

The S=1 chain has a quite different physics~: the isotropic 
antiferromagnetic chain
is in the Haldane phase. There is a gap for spin excitations and the
correlations decay exponentially. When adding anisotropy, this
Haldane phase survives in a region of parameters around the isotropic
point\cite{schulz,DNR}. When anisotropies become too large, then the
system is eventually driven towards more conventional ferro or
antiferro phases\cite{schulz}.

This spin-1 chain is closely related to the two-leg spin-1/2 ladder.
In the ladder system, one is dealing with two S=1/2 spin chains
coupled by rungs perpendicular to the chains. The ladder has a gap
for all values of the perpendicular coupling. In fact, it is known
that the S=1 chain and the two-leg spin ladder share the same
low-energy effective field theory\cite{SM,STN}.

Understanding the influence of disorder on such spin systems is an
important issue. In all real materials there is of course some amount
of disorder. One of the probe of the physics of the system is to
introduce artificially defects or impurities.
 The study of disorder in one-dimensional quantum
systems is also of considerable theoretical interest\cite{Fisher}.
Recently it has been realized that it is possible to obtain exact
results in such disordered systems, while this is generally
impossible in the classical realm.

In the case of the spin-1/2 chain, one simple kind of disorder is a
distribution of quenched bond disorder. The real-space
renormalization group\cite{MDH} study shows that the ground state is
a collection of singlet bonds over arbitrarily large distances in a
random pattern according to the initial distribution of randomness.
This is the so-called random singlet phase\cite{BhattLee}. The
influence of weak disorder on the spin-1/2 chain has been also
studied by means of a renormalization group (RG) study on the
effective field theory\cite{DotyFisher}. This method is a
perturbative RG calculation but it is easy to study widely different
kinds of disorder. The real-space RG method has also been applied to
systems with a distribution of ferro- and antiferromagnetic
bonds\cite{West} as well as dimerized systems\cite{Ha,Hb}. There are
also direct numerical studies\cite{Hida}.

In the case of the spin-1 chain we are
dealing with a gapped system that has also some hidden long-range
order as exhibited by the approximate valence-bond-solid (VBS) 
ground state
wavefunctions\cite{AKLT,DNR}. Certainly one expects the influence of
disorder to be quite different\cite{Boe}. A real-space RG has been
performed on the isotropic S=1 chain\cite{HY}. It shows that there is
destruction of the Haldane gap beyond a nonzero disorder strength,
then there is a gapless phase which sustains the hidden long-range
order and then for even stronger disorder one finds the featureless
random-singlet phase. This is analogous to the doping of conventional
superconductors with magnetic impurities\cite{MGJ}~: here the hidden
long-range order plays the role of the superconducting condensate.

In this paper we study the S=1 spin chain with exchange and
single-ion anisotropies.
This model has a rich phase diagram and the influence
of disorder is extremely sensitive to the nature of the ground state
of the unperturbed system.
 We use a
perturbative RG treatment of the disorder on the effective field
theory. The average over the disorder
is performed by use of the replica trick as pioneered by Giamarchi
and Schulz\cite{Giam} in the context of the one-dimensional electron
gas. Here we perform the calculation up to second order in the
disorder strength. This is important as shown by Fujimoto and
Kawakami\cite{Fuji} because, at this order of perturbation, one
captures the interplay between interactions and disorder when they
are both relevant. In the spin-1/2 case they have shown that it is
possible to explicitly obtain from the RG treatment the destruction
of the antiferromagnetic phase under a random z-axis magnetic field
for an arbitrarily small amount of randomness as predicted by the
argument of Imry and Ma\cite{ImryMa}. The case of the spin 
ladder\cite{Ori} is closely related to our work but the relevant
operators are not always the same. In the spin-1 chain
there are operators generated by disorder that involve
products of the two spin-1/2 operators used to describe
the spin-1 degrees of freedom. They do not exist in the ladder case.

In section II, we give the construction of the effective bosonic
theory for anisotropic S=1 chain without randomness. In section III,
we translate in the bosonic language, the operators induced by the
disorder. Section IV shows why gapped phases are in general stable
under weak disorder. Section V contains the derivation of the
renormalization group equations. Section VI gives the results of our
study on each phase of the S=1 chain.  Finally section VII contains 
our conclusions.

\section{The pure S=1 spin chain}

In this section we give the construction of the effective
Bosonic low-energy theory for the S=1 spin chain. 
We will use this continuum theory as a starting point
for the treatment of randomness in sect. III.
We concentrate on the anisotropic Heisenberg S=1
antiferromagnetic spin chain with the following Hamiltonian~:
\begin{equation} 
{\mathcal H} = \sum_i S_i^x S_{i+1}^x + S_i^y
S_{i+1}^y + \Delta S_i^z S_{i+1}^z + D(S_i^z)^2.
 \label{S1}
\end{equation} 
The $S_i^\alpha$ are S=1 quantum spin operators,
$\Delta$ is the exchange anisotropy and $D$ is the single-ion
anisotropy. All energies are measured in units of $J$.
The phase diagram of this model is well-known
in the absence of disorder\cite{DNR}. There are two phases
with conventional long-range magnetic order~:
a ferromagnetic (F) phase for $\Delta$ large and negative and an
antiferromagnetic (AF) phase for $\Delta$ large and positive.
In the intermediate regime, there is the Haldane phase (H)
which is gapped and has no obvious long-range order.
In particular it does not break translational or rotational
symmetry. There are two gapless XY-like phases that differ
only in the behaviour of the spin correlations. Finally,
when $D$ is very large and positive there is a non-magnetic phase which
is trivially obtained by perturbation theory from the strong-coupling
limit where the ground state has all spins set to $S_{i}^{z}=0$~:
$|000\dots 00\rangle$, called the large-D phase (D) in what follows.

The starting point is to write each spin-1 operator as the sum of two
spin-1/2 operators $S_i^\alpha = \sigma_i^\alpha + \tau_i^\alpha$,
an approximation suggested by Luther and Peschel\cite{Luther}. 
It has been shown by Schulz that Abelian
bosonization\cite{schulz} is able to reproduce 
most of the phase diagram.
One first performs a Jordan-Wigner transformation on each
spin-1/2 $\bf \sigma$ and $\bf \tau$ by
introducing two species of spinless fermions. 
Explicitly we write~:
\begin{mathletters} 
\begin{eqnarray} 
&& 
\sigma_n^+ = (-)^n
c_n^{\dagger} {\rm e}^{i\pi\sum_{m<n}c_m^{\dagger}c_m} , \\ 
&&
\sigma_n^- =  (-)^n c_n {\rm e}^{i\pi\sum_{m<n}c_m^{\dagger}c_m} , \\
&& 
\sigma_n^z = c_n^{\dagger}c_n - \frac{1}{2}\,\, , 
\end{eqnarray}
\end{mathletters} 
where $c$ is a fermion operator and similarly one
introduces a $d$ fermion for the other set of spins-1/2
$\tau_n^\alpha$. Since one is interested in the long-wavelength, 
low-energy behaviour, it is then convenient to take the continuum limit
by introducing two continuous Fermi fields $\psi_{c} (x)$ and
$\psi_{d} (x)$. The Hamiltonian (\ref{S1}) leads then to a theory
of two interacting Fermi fields in one space dimension.
The physical content of this Fermi system
is obtained by translation into the language of 
interacting Bosons\cite{Boso}. In fact, in one space dimension
a theory involving a Fermi field $\psi (x)$ can be
translated in a Bose field theory by the following relations~:
\begin{mathletters} 
\begin{eqnarray} 
&& \psi_L = \frac{1}{\sqrt{2\pi
\alpha}}e^{i\sqrt{4\pi}\phi_L} , \\ && \psi_R = \frac{i}{\sqrt{2\pi
\alpha}}e^{-i\sqrt{4\pi}\phi_R} , 
\end{eqnarray} 
\end{mathletters}
where $\phi_L$ and $\phi_R$ are the chiral components of a Bose
field~: $\phi_{L,R}= 1/2(\phi \mp \int^x \Pi (x^\prime
)dx^\prime )$. Thus there are two Bose fields $\phi_c$ and $\phi_d$
corresponding to the two fermions $\psi_{c}$ and $\psi_{d}$. 
The effective theory
for the spin-1 chain can be simplified by use of the natural
``acoustic'' and ``optic'' linear combinations~: 
\begin{equation}
\phi_a = {1\over \sqrt{2}}(\phi_c + \phi_d ) \quad {\rm and} \quad
\phi_o = {1\over \sqrt{2}}(\phi_c - \phi_d ) . 
\end{equation} 
The
Bose theory can then be written as~: 
\begin{equation} 
{\mathcal H} =
{\mathcal H}_a + {\mathcal H}_o , 
\label{Btheo}
\end{equation} 
where the a-sector
is a sine-Gordon theory~: 
\begin{equation} 
{\mathcal H}_a =
\frac{v_a}{2} \left [ K_a (\Pi_a^2) + \frac{1}{K_a} (\nabla \phi_a)^2
\right ] + \frac{g_1}{(\pi \alpha)^2} \cos (\sqrt{8 \pi}\phi_a) ,
\label{SG} 
\end{equation} 
and the o-sector is a {\it generalized}
sine-Gordon theory~: 
\begin{equation} 
{\mathcal H}_o =\frac{v_o}{2}
\left [ K_o (\Pi_o^2) + \frac{1}{K_o} (\nabla \phi_o)^2 \right ]
+\frac{g_2}{(\pi \alpha)^2} \cos (\sqrt{8 \pi}\phi_o) +
\frac{g_3}{(\pi \alpha)^2} \cos (\sqrt{2 \pi} \tilde{\phi}_o).
\label{GSG} 
\end{equation} 
Here $\tilde{\phi}_o = \phi_{Lo}
-\phi_{Ro}$ is the dual field of $\phi_o$, and the couplings are
given by $g_1 = g_2 = D-\Delta$, $g_3 = -1$ at first order in 
$D, \Delta$. The velocities $v_a$, $v_o$ and the two parameters $K_a$,
$K_o$ are also functions of the initial parameters of the problem and
can be computed to the first order in $D$ and $\Delta$~:
\begin{equation}
v_a = 1+ {3\Delta +D\over \pi},\quad
v_o = 1+ {\Delta -D\over \pi},\quad
K_a = 1-  {3\Delta +D\over \pi},\quad
K_o = 1- {\Delta -D\over \pi}.
\label{first}
\end{equation}
Contrary to the case of the S=1/2 chain, there are no exact results
on the values of the couplings  $K_a$,  $K_o$ as functions of the 
bare parameters $D$ and $\Delta$. Strictly speaking, there are 
operators coupling the two fields $\phi_a$ and $\phi_o$ in 
Eq.(\ref{Btheo}) but they are less relevant than the interactions
in Eq.(\ref{SG}) and Eq.(\ref{GSG}).

The physics of
the a-sector is well-known~: there is a Kosterlitz-Thouless phase
transition separating a massless phase with algebraic quasi-long
range order from a massive phase in which the field $\phi_a$ has a
definite vacuum expectation value. The transition takes place when
the operator $\cos (\sqrt{8 \pi}\phi_a)$ becomes relevant according
to the value of the parameter $K_a$~:
this happens for $K_{a}=1$. In the massive regime $K_{a}<1$
the field
$\phi_a$ acquires a vacuum expectation value $\langle\phi_a\rangle$
fixed by the minimum of the cosine operator appearing in the energy.
In the massless regime, correlations functions decay algebraically
with powers dictated by the value of the coupling $K_{a}$.

The physics of the o-sector is
less transparent~: it involves now the dual field $\tilde{\phi}_o$
which is a non-local operator as a function of $\phi_{o}$. In
the Coulomb gas language, this theory has both electric and magnetic
charges. In fact, den Nijs\cite{dN} has shown that this effective
theory appears also in the Coulomb gas representation of the XY model
with Ising anisotropy. Since this model undergoes an {\it Ising}
phase transition between two massive phases, we expect the same to be
true for the generalized sine-Gordon theory (\ref{GSG}).

The different phases of the S=1 spin chain can be characterized from
the behavior of the effective theories (\ref{SG}) and (\ref{GSG}).
In the AF phase, both fields $\phi_a$ and $\phi_o$ have a vacuum 
expectation value. In the Haldane and
large-D phase it is now $\phi_a$ and $\tilde{\phi_o}$ that have
expectation values. It is the Ising transition of the theory
(\ref{GSG}) that takes place
between AF and Haldane and also between AF and large-D. 
The difference between
the Haldane and large-D phases is only the sign in front of the 
operator 
$\cos (\sqrt{8 \pi}\phi_a)$. Thus it is the value of the condensate
$\langle \phi_a \rangle$
that differs between the Haldane and large-D phases. When the 
coefficient
of the operator $\cos (\sqrt{8 \pi}\phi_a)$ is zero, there is a
massless line in the phase diagram separating H and D. 
When $K_{a}$ is large enough, the field $\phi_a$ is massless~:
when  $\phi_o$ condenses it the XY2 phase while $\tilde{\phi_o}$
condenses in XY1 phase. These characteristics are
summarized in table~1 (where ``dis'' means that the corresponding
field has no mean value and is disordered).
Finally, there is a
ferromagnetic phase (F) where bosonization breaks down~: at the
F boundary the velocity $v_a$ vanishes. The F phase
is beyond the scope of our study and will not be discussed.

The phase boundaries are in good agreement with the more detailed
picture of den Nijs and Rommelse\cite{DNR}. It should be noted that
the first-order expressions for $K_{a}$ and $K_{o}$ 
in Eqs.(\ref{first}) lead to 
straight lines for the phase boundaries while in reality
there are tricritical points, i.e. between Haldane, large-D and AF
phases. They are also out of reach of the bosonization method
at the order which is considered here.

\section{Disorder in the bosonic language}

In this section we describe an approximate treatment of quenched 
randomness
in the Bosonized theory obtained in sect. II. We give the operators 
that should be added to the Bosonic field theory (\ref{Btheo})
to describe randomness. We first list the various kinds of randomness
that we have been able to treat.
We will consider the effects of a random 
field along the z-axis. We add to the Hamiltonian
a term~:
\begin{equation} 
{\mathcal H}_{ZF}=\sum_{i} h_{i} S^z_{i} ,
\label{zf}
\end{equation} 
where the local fields $h_{i}$ are Gaussian random variables
with zero mean,
uncorrelated from site to site~:
\begin{equation}
\overline{h_{i} h_{j}} = {\mathcal D}_{ZF} \delta_{ij}.
\end{equation}
The bar means average over the random distribution and the variance
of the distribution is ${\mathcal D}_{ZF}$, characterizing the 
strength of
the disorder.
This kind of disorder preserves the rotational symmetry around the 
z-axis which a symmetry of the spin hamiltonian.
It is known from the Imry-Ma argument\cite{ImryMa} that it has a 
dramatic 
effect
on the phases with long-range magnetic order~: an arbitrarily small 
amount of random z-field destroy the ordering.
We consider also the case of a random planar field, i.e. lying 
in the XY plane~:
\begin{equation} 
{\mathcal H}_{PF}= \sum_i h_i^x S_i^x + h_i^y S_i^y ,
\label{pf}
\end{equation} 
with 
$ 
\overline{h_{i}^{\alpha} h_{j}^{\beta}} = {\mathcal D}_{PF} 
\delta_{ij}
\delta_{\alpha\beta} .
$
The rotational symmetry is destroyed in this case.
We treat  random planar exchange~:
\begin{equation}
	{\mathcal H}_{PE}=\sum_i J_i ( S_i^x S_{i+1}^x + S_i^y S_{i+1}^y), 
	\label{pe}
\end{equation}
with 
$\overline{J_{i}J_{j}}=  {\mathcal D}_{PE} \delta_{ij}$.
Finally randomness may break XY-symmetry by random planar anisotropy~:
\begin{equation}
	{\mathcal H}_{PA}=\sum_i \gamma_i ( S_i^x S_{i+1}^x - 
S_i^y S_{i+1}^y),
	\label{pa}
\end{equation}
with 
$\overline{\gamma_{i}\gamma_{j}}=  {\mathcal D}_{PA} \delta_{ij}$.

When bosonizing the spin operators in 
Eqs.~(\ref{zf},\ref{pf},\ref{pe},\ref{pa}),
there is one part with small momenta $q\approx 0$ and one part
with $q\approx\pi$. For example, in the ZF case we have~:
\begin{equation}
	S^{z}(x, t)
	= -\sqrt{\frac{2}{\pi}} \partial_x
    \phi_a +  \frac{(-)^{x}}{\pi \alpha}\cos(\sqrt{2
    \pi}\phi_a) \cos(\sqrt{2 \pi}\phi_o),
	\label{SZ}
\end{equation}
in the continuum limit. Inserted in Eq.(\ref{zf}) this formula
will pick the $q\approx0$ part of the fluctuating field $h_{i}$
as well as the $q\approx\pi$ part~:
\begin{equation}
	{\mathcal H}_{ZF}= \int dx \,
	 h^{q\approx0}(x)\, {\mathcal O}_{1} (x, t)
	+  h^{q\approx\pi}(x)\, {\mathcal O}_{2} (x, t)	,
	\label{bzf}
\end{equation}
with~:
\begin{equation} 
{\mathcal O}_{1}=-\sqrt{\frac{2}{\pi}}\partial_x \phi_a (x, t)
\quad
{\rm and}
\quad
{\mathcal O}_{2}=\cos(\sqrt{2
    \pi}\phi_a) \cos(\sqrt{2 \pi}\phi_o) .
\end{equation}
In fact this formula is generic for all kinds of disorder. We have 
listed in Table II the operators ${\mathcal O}_{1}$ and 
${\mathcal O}_{2}$
corresponding to all cases cited above. 
We have kept only the most relevant operators in the
renormalization group sense.
The continuous random fields
$h^{q\approx 0}(x)$ and $h^{q\approx\pi}(x)$ are also Gaussian.
To treat the quenched disorder we follow the replica trick~:
we introduce $n$ copies of the system and compute the free energy
as 
$F=\lim_{n\rightarrow 0} {\overline{(Z^{n}-1)}}/n$. Integrating
over the Gaussian disorder $h^{q\approx0}(x)$
leads
in the Euclidean effective action to the following operator~:
\begin{equation} 
{{\rm S}_{\rm eff}}=
-{{\mathcal D}_{1} }
\int dx \, d\tau \, d\tau^{\prime}
 \, \sum_{i,j} \, {\mathcal O}_{1}^{(i)}(x, \tau)
 {\mathcal O}_{1}^{(j)}(x, \tau^{\prime}),
\label{diso1}
\end{equation}
where ${\mathcal D}_{1}$ is the variance of $h^{q\approx0}(x)$,
$i, j$ are replica indices. It is important to note
that this operator is not local~: since the random field 
$h^{q\approx0}(x)$ is quenched i.e. does not depend on time,
the Gaussian integration leads to the two-time integral
in Eq.(\ref{diso1}). 
We obtain exactly similar formulas for all operators
given in table II. In each case there is a variance $\mathcal D$
corresponding to the Gaussian random field coupled to the operator 
considered. We will use the obvious notation $ZF^{(0)}$, 
$ZF^{(\pi)}$, $PF^{(0)}$, and so on. The disorder
strengths are noted also ${\mathcal D}_{ZF}^{(0)}$,
 ${\mathcal D}_{ZF}^{(\pi)}$, \dots .
The $ZF^{(0)}$ operator corresponds to a forward scattering process
in the language of the Jordan-Wigner fermions. When written
as a Bosonic operator it may be absorbed by a change of variable
in the path integral over the field $\phi_a$ which leads to
simple leading-order RG calculations\cite{Giam},
when interactions in the a-sector are irrelevant.
Here we will retain this $ZF^{(0)}$ operator explicitly
in our RG treatment because, when its strength diverges,
it leads to a very important phenomenon~: the vanishing
of the critical disorder strength for massive phases.

It is important to note that the most relevant $q=\pi$ operators
are, in all cases over study, always given by $ZF^{(\pi)}$.
This is due to the fact that even when $ZF^{(\pi)}$ does not
appear at the bare level, it is generated by renormalization~:
this has been pointed out first by Doty and Fisher in the study
of the S=1/2 chain\cite{DotyFisher}.
In our case a similar reasoning leads immediately to the 
appearance of $ZF^{(\pi)}$ in the PF case as well as in the PA
case (in the ZF and PE case, $ZF^{(\pi)}$ appear at the bare level)

The operators  $PA^{(0)}$ and $PE^{(0)}$
are specific to the S=1 spin chain as opposed to the spin ladder~:
they come from the cross coupling 
${\sigma}_{i}\cdot{ \tau}_{i+1}$
that occurs
when expressing S=1 spin operators into S=1/2 entities.
In the spin ladder it would be replaced by 
${\bf\sigma}_{i}\cdot{\bf \tau}_{i}$
across the rung.

The operators  of the type (\ref{diso1}) break the Lorentz invariance
which is present in the effective field theory of the pure system.
This means that in a renormalization group calculation many
new operators will be generated in addition to those
introduced at the bare level. In principle it is thus extremely
difficult to keep the RG flow under control.
In the RG calculation at second order that we performed, we have
to add for internal RG consistency only three new operators~:
\begin{mathletters} 
\begin{eqnarray} 
&& 
{\mathcal S}_{a1} =
 -{\mathcal D}_{a1} \int dx \, d\tau \, d\tau^{\prime} \,
\sum_{i,j} \, \partial_\tau \phi^i_a(x,\tau ) \partial_\tau
\phi^j_a(x,\tau^{\prime}) , \\ 
&&
{\mathcal S}_{a2}= 
-{\mathcal D}^{}_{a2} \int dx \, d\tau \,
d\tau^{\prime} \, \sum_{i,j} \, \partial_w \phi^i_a(x, \tau )
\partial_{\overline{w}} \phi^j_a(x, \tau^{\prime}) , \\
&& 
{\mathcal S}_{o}=
 -{\mathcal D}^{}_{o}
\int dx \, d\tau \, d\tau^{\prime} \, \sum_{i,j} \, \partial_z
\phi^i_o(x,\tau ) \partial_{\overline{z}} 
\phi^j_o(x,\tau^{\prime}) , 
\end{eqnarray}
\end{mathletters} 
where
$z=x+ v_o \tau$,
$\overline{z}= x-v_o \tau$,
$w=x+v_a \tau$ and $\overline{w}= x-v_a \tau$. They are called 
$a1$, $a2$ and $o$ in what follows.

\section{renormalization group treatment}

In this section we perform a second-order RG treatment of the Bosonic 
theory given by the field theory Eq.\ (\ref{Btheo}) perturbed by 
the 
operators of the type (\ref{diso1}). There are five distinct bare
operators ($ZF^{(0)}$, $ZF^{(\pi)}$, $PF^{(0)}$, $PE^{(0)}$, $PA^{(0)}$)
and three generated by renormalization ($a1$, $a2$, $o$).
The perturbation theory can start only from the massless theory
with no cosine operators $g_{1} = g_{2} =g_{3}=0$ in Eq.(\ref{Btheo}).
We write the effective theory including randomness as~:
\begin{equation} 
S=S^* + \sum_i \, a^{x_i-2} g_i \int dx\, d\tau\, {\mathcal O}_i ,
\end{equation} 
where $S^{*}$ is the massless theory characterized by $K_{a}$ and
$K_{o}$, $a$ is the cut-off and $x_{i}$ is the scaling dimension
of the operator  ${\mathcal O}_i$.
The beta functions of the couplings $g_{i}$ are then given
by the following formula~:
\begin{equation} 
\frac{dg_k}{dl} =
(2-x_k)g_k - \pi \sum_{ij} C_{ijk} \, g_i g_j + O(g^3),
\end{equation} 
where the Wilson coefficients $C_{ijk}$ are obtained from
the Operator Product Expansion~: 
\begin{equation} 
{\mathcal O}_i (z).{\mathcal
O}_j (0) \sim \sum_k \frac{1}{|z|^{x_k -x_i - x_j}}C_{ijk} 
{\mathcal O}_k(0). 
\end{equation} 
This procedure gives the beta function for an
arbitrary number of replicas and is very useful in order to find
which operators are generated by renormalization. 
While this technique can be applied straightforwardly to
local operators like those appearing in the sine-Gordon
or generalized sine-Gordon theories, this is not so when dealing
with the non-local operators (\ref{diso1}) generated by randomness.
We will take into account 
 only the strongest singularities (which are in $1/|Re \,
z|$) in the OPE containing nonlocal operators, and assume that the 
OPE is still valid.

We now sum up the renormalization group equations of the system.
In the replica $n\rightarrow 0$ limit, we obtain the flow for 
the cosine couplings~: 
\begin{mathletters}
\label{ys}
\begin{equation} 
\frac{dy_1}{dl} =
 2(1-K_a) y_1 
 - \frac{1}{8\pi^2}K^2_a y_1
( 2{\mathcal D}^{(0)}_{ZF} + 2{\mathcal D}_{a1} 
+ {\mathcal D}_{a2}) , 
\label{y1} 
\end{equation}
\begin{equation}
\frac{dy_2}{dl} =
 2(1-K_o)y_2 - \frac{1}{8 \pi^2}K^2_o {\mathcal D}_{o} y_2 , 
\label{y2}
\end{equation}
\begin{equation}
\frac{dy_3}{dl} =
 (2-\frac{1}{2K_o})y_3 - \frac{1}{32 \pi^2} {\mathcal D}_{o} y_3 , 
\label{y3} 
\end{equation} 
\end{mathletters} 
where 
$y_1 = g_1/v_a$, $y_{2,3} = g_{2,3}/v_o$.
In the first equation (\ref{y1}), in the absence of disorder 
the renormalization eigenvalue $2(1-K_a)$ leads to a massive
phase when $ K_a < 1$~: the cosine is relevant since
$y_1$ scales to infinity in this case. This picture may be 
altered since the randomness leads to an additional {\it 
second-order}
coupling
$ - y_1\times {\mathcal D}^{(0)}_{ZF}$ between the strength of the 
interaction and the forward scattering term $ZF^{(0)}$. If 
${\mathcal D}^{(0)}_{ZF}(l)$ scales fast enough to infinity then 
it can 
revert the flow of $y_1$ and leads to the destruction of the 
massive
phase. This behaviour has been obtained in the case of the Hubbard
model by Fujimoto and Kawakami\cite{Fuji}. We will show later 
that this 
happens also in the case of the spin chain under study.
Of course to capture such phenomena, it is important to deal
explicitly with $ZF^{(0)}$ instead of eliminating it by 
a change of variables in the functional integral.

There are also two flow equations for the stiffness constants~:
\begin{mathletters}
\label{ks}
\begin{equation} 
\frac{dK_a}{dl} =
 -( {\mathcal D}_{ZF}^{(\pi)} + \frac{2}{\pi^2} y_1^2) K_a^2
+\frac{1}{4}{\mathcal D}_{PF}^{(0)} + {\mathcal D}_{PA}^{(0)},
\label{ka}
\end{equation}
\begin{equation}
 \frac{dK_o}{dl} =
  -(\frac{{\mathcal D}_{ZF}^{(\pi)}}{u} +
\frac{2}{\pi^2}y_2^2)K_o^2 
+(\frac{{\mathcal D}_{PF}^{(0)}}{4u} +
\frac{{\mathcal D}_{PE}^{(0)}}{u} + \frac{1}{\pi^2}y_3^2) ,
\label{ko}
\end{equation} 
\end{mathletters} 
where $u = v_o/v_a$.
The operators that do not involve gradients have simple
renormalization properties. In fact they scale according to the
dimension that can be found immediately from the massless theory~:
\begin{mathletters}
\label{dees}
\begin{equation} 
\frac{d{\mathcal D}_{ZF}^{(\pi)}}{dl} =
 (3- K_a - K_o ){\mathcal D}_{ZF}^{(\pi)} , 
\label{de1}
\end{equation}
\begin{equation} 
 \frac{d{\mathcal D}_{PF}^{(0)}}{dl} = (3- \frac{1}{4K_a} -
\frac{1}{4K_o}){\mathcal D}_{PF}^{(0)} ,  
\label{de2}
\end{equation}
\begin{equation} 
 \frac{d{\mathcal D}_{PE}^{(0)}}{dl} =
 (3 - \frac{1}{K_o}){\mathcal D}_{PE}^{(0)} , 
\label{de3}
\end{equation}
\begin{equation} 
  \frac{d{\mathcal D}_{PA}^{(0)}}{dl} =
 (3- \frac{1}{K_a}){\mathcal D}_{PA}^{(0)} .
\label{de4}
\end{equation} 
\end{mathletters} 
These equations allow immediate statements about the stability
of the massless phases according to the relevance or irrelevance
of the random operators.
This simplicity however does not extend to the operators
$ZF^{(0)}$, $a1$, $a2$ and $o$ because they appear in the OPE
of various combinations of the other operators. The RG equations
thus couple the various random perturbations~:
\begin{mathletters} 
\label{moredees}
\begin{equation} 
 \frac{d{\mathcal D}_{ZF}^{(0)}}{dl} =
{\mathcal D}_{ZF}^{(0)} 
+ \pi (\frac{1}{4}f(u){\mathcal D}_{ZF}^{(\pi)\, 2} -
\frac{1}{8}g(u){\mathcal D}_{PF}^{(0)\, 2} 
- \frac{1}{2}h {\mathcal
D}_{PA}^{(0)\, 2}) , 
\label{cc} 
\end{equation}
\begin{equation}
\frac{d{\mathcal D}_{o}}{dl} =
 {\mathcal D}_{o} + \pi
(\frac{1}{4u}f(u){\mathcal D}_{ZF}^{(\pi)\, 2} 
- \frac{1}{8u}g(u){\mathcal D}_{PF}^{(0)\, 2}
- \frac{1}{2u} k(u) {\mathcal D}_{PE}^{(0)\, 2}) ,
\label{ccc} 
\end{equation} 
\end{mathletters} 
Equations similar to
(\ref{moredees}) hold also for the couplings 
${\mathcal D}_{a1}$ and ${\mathcal D}_{a2}$. 
The second-order terms in the 
flow equations above involve the following functions~:
\begin{mathletters} 
\begin{equation} 
f(u) = \left [
\int^{+\infty}_{-\infty} \frac{dy}{(1+y^2)^{K_a/2} (1+u^2
y^2)^{K_o/2}} \right ]^2 , 
\quad
g(u) =
\left [ \int^{+\infty}_{-\infty} \frac{dy}{(1+y^2)^{1/8K_a} (1+u^2
y^2)^{1/8K_o}} \right]^2 , 
\label{functions1}
\end{equation} 
\begin{equation} 
h =
\left [ \int^{+\infty}_{-\infty} \frac{dy}{(1+y^2)^{1/2K_a}}
\right]^2 , 
\quad
k(u) = \left [
\int^{+\infty}_{-\infty} \frac{dy}{(1+u^2 y^2)^{1/2K_o}} \right]^2 .
\end{equation} 
\end{mathletters} 
It is important to note that these functions are
not well-defined for all values of the stiffness constants $K_a$,
$K_o$. We will show in the following section that these divergences
lead to the disappearance of the Ising-ordered phase for an
arbitrarily weak random z-field, as predicted by the Imry-Ma 
argument.

Armed with the flow equations 
(\ref{ys},\ref{ks},\ref{dees},\ref{moredees}), 
we can study the phase diagram
of the S=1 chain in the presence of the various kinds of randomness
listed in Table II. Before embarking in the general case, we briefly
discuss a simplified model that display most of the behaviour
of the complete theory~:
\begin{equation} 
{\mathcal S}_{eff} = \frac{1}{2K} \int \, dx \, d\tau
(\partial_{\mu} \phi)^2 + \frac{g}{(\pi \alpha)^2} \int \, dx \, 
d\tau \cos(\sqrt{4 \pi a^2 }\phi(x, \tau)) - 
\nonumber 
\end{equation}
\begin{equation} 
-\frac{d}{(\pi \alpha)^2} \int \, dx \, d\tau \,
d\tau^{\prime} \cos ( \sqrt{4 \pi b^2} \tilde{\phi}(x, \tau)) 
\, \cos (\sqrt{4 \pi b^2}\tilde{\phi}(x, \tau^{\prime})). 
\end{equation} 
Defining $y=g/\pi u$, and $D=d \alpha / 2\pi u^2 $, where 
u is the velocity, we derive  the following renormalization group
equations for y, D and K~: 
\begin{mathletters} 
\label{reno}
\begin{eqnarray} 
&&
\frac{dy}{dl}= (2-a^2 K)y , 
\\ 
&& 
\frac{dD}{dl}= (3-b^2 /K)D , 
\\ 
&&
\frac{dK}{dl}=-y^2 K^2 + D . 
\end{eqnarray} 
\end{mathletters} 
The
corresponding RG flow is given in Figure~2 when $(ab)^2 < 6$. 
Here, the crucial parameter is the stiffness constant K, because it
governs the relative relevance of interaction and disorder operators.
For instance, if K flows to zero for small initial disorder, the
disorder operator will become irrelevant, whereas the interaction
operator is relevant. If the pure case corresponds to a regime where
interactions are relevant, the presence of infinitesimally small (but
finite) disorder will not affect this behavior. The value of $K^*$ in
the pure case is enough to predict the qualitative behavior of the
system under weak disorder at this order of perturbation theory.
It is clear that the
massive phase will persist up to some critical disorder which,
strictly speaking, is outside of reach of perturbation theory. A
formal proof of stability is given in appendix A.
The flow equations (\ref{reno}) are only first-order. The interest
of our second-order study will appear in the study of the 
Ising-ordered phase.

Finally, we comment on the correctness of the bosonization approach.
Since the scheme we use is perturbation improved by use
of the renormalization group then we can't say much
about the flow to strong coupling. The phases that are unstable
cannot be characterized by our tools. We can only find
whether or not the randomness will have an immediate effect
on the system. Even in the stable case, we of course expect
that for strong enough disorder the physics of the system will
change through a phase transition of some kind. This strategy
has been applied with some success in the S=1/2 chain
in a random z-field
ans in the related problem of 1D fermions with attractive 
interactions
submitted to a random site potential. 
Bosonization\cite{Giam,DotyFisher} predicts
stability of the XY phase of the S=1/2 chain in the region
$-1<\Delta < -1/2$ and there is ample evidence for the correctness
of this result\cite{Pang,Haas,Runge,Eckern} from direct numerical 
studies.
Our general study should be followed by attacks by similar methods
for an independent consistency check.

\section{Phase diagram in the presence of disorder}
\label{phasediag}
In this section, we exploit the RG flow equations obtained previously 
to discuss the influence of randomness on the phase
diagram of the S=1 spin chain phase diagram pictured in Fig. (1).
We begin our discussion by the massive phases, with and without
long-range spin order. All our findings are summarized in Table III.
The key element of the stability is the operator $ZF^{(\pi)}$
since it is present in all kinds of randomness.

\subsection{AF phase}

This phase is characterized by the following IR stiffness constants
in the pure case : $K_a^*=K_o^*=0$. 
For such a fixed point, the functions appearing at second-order
in Eqs.\ (\ref{moredees}) take the values $g(u) = h = k(u) =
0$ while the function $f$ (\ref{functions1}) is singular. 
The $ZF^{(\pi)}$ operator is {\it relevant} in this phase.
Since the cosine interaction is relevant in the a-sector,
it seems that we are in a case with interaction and disorder
that are relevant. However, the second-order beta functions
allow a more precise conclusion~: when $K_a^*$ and $K_o^*$
flow to a small enough value then $f$ diverges. When approaching
this singular point, ${\mathcal D}_{ZF}^{(0)}$ becomes
arbitrarily large even for infinitesimally small values
of ${\mathcal D}_{ZF}^{(\pi)}(l=0)$ and 
${\mathcal D}_{ZF}^{(0)}(l=0)$. In Eq. (\ref{y1}), the coupling
$ - y_1\times {\mathcal D}^{(0)}_{ZF}$ becomes so large that
the interaction $y_1$ is driven to zero. We infer thus that the
Mott-like gap of the Ising phase is immediately destroyed
by disorder, i.e. by infinitesimal disorder.

In the case of a random z-field, this is really what we predict
from the Imry-Ma reasoning. In the case of the random planar
field and random planar anisotropy, the random z-field
is generated by the RG process so this is why there is also
destruction of the Ising phase. The only trouble with the 
bosonization
treatment is that it also leads to instability under
random planar exchange, a fact which is not expected from the
Imry-Ma argument. Since stability in this case is expected 
we see this as a problem of our approach
(if we apply the same treatment to the S=1/2 case then
one finds also instability under random planar exchange
of the corresponding Ising phase). However
this approach correctly reproduces the immediate destruction
of the gap in the Ising phase.

\subsection{Haldane and large-D phases} 

In the renormalization group
equations, nothing distinguishes the H and D phases. So we expect
that they behave in the same way under disorder. These phases are
characterized by $K_a^* = 0$, $K_o^* = \infty$, and $\phi_a$ has a
vacuum expectation value. 
The function $f$ is now well defined
when we reach the unperturbed fixed point, and near this fixed point,
$f(u) = 0$. The operators $ZF^{(\pi)}$, $PF^{(0)}$ and $PA^{(0)}$
are {\it irrelevant}. This implies immediately
stability under random XY symmetry-breaking exchange.
Concerning the random fields, we note that
the random forward scattering ${\mathcal D}_{ZF}^{(0)}$
scales as ${\rm e}^l$ while $y_1$ and $y_3$ scale faster as 
${\rm e}^{2l}$
according to Eqs.~(\ref{cc},\ref{y1},\ref{y3}).
Thus there is no possibility of
destruction of the role of interactions contrary to the case of
the AF phase. Thus there is stability under random z-field
as well as under random XY-field.

The only special case is the random XY-symmetric coupling.
The $PE^{(0)}$ operator is relevant
and the function k(u) is singular. This drives the $o$-operator
to large negative values of ${\mathcal D}_{o}$. 
This goes outside the reach of perturbation theory.
But for such large
negative values, $y_3$ scales much faster than the disorder like
$e^{|D_{o}|l}$. We take this as an indication of the robustness
of the gap against disorder. So we expect stability also
in this case.
The two gapped phases without long-range spin order are thus
stable, up to some critical strength of the disorder presumably,
which is beyond the reach of the methods we employ here.

\subsection{XY1 phase}

In this phase, the a-sector is gapless, and
the $ZF^{(\pi)}$ operator is  irrelevant near the 
unperturbed fixed point in the whole phase so instabilities
can appear only through the other operators induced by randomness.

\noindent 
$\bullet$ Random z-field~:  
the forward scattering $ZF^{(0)}$ operator can be absorbed
by a field redefinition. Thus the excitation spectrum 
is unchanged and the correlation functions are affected 
in a simple way\cite{Giam}.
\bigskip

\noindent 
$\bullet$ Random XY field~: the $PF^{(0)}$ operator is relevant; 
so the phase is unstable.
In this case, we note that the function 
$g$ is singular, so the $ZF^{(0)}$ operator is singular too and 
scale to $+ \infty$.
\bigskip

\noindent 
$\bullet$ 
Random XY symmetric coupling~: the $PE^{(0)}$ operator is relevant,
as in the Haldane/large-D phase. The same discussion apply~: the
stability of the o-sector lead us to conclude that the XY1 phase
remains stable under this kind of disorder. 
\bigskip

\noindent 
$\bullet$ 
Random XY symmetry-breaking coupling~: 
The $PA^{(0)}$ operator is relevant and the $a$-sector is not
gapped; so the phase is unstable.
The operator leading to the instability is different from 
the ladder case because it involves $\sigma\tau$ couplings
that are typical of the S=1 case.

\subsection{Frontier line between Haldane and large-D phases}

This line is characterized by $y_1 = 0$, so that the field $\phi_a$
is a free massless field. There is no gap along this line in the
$a$-sector. It can be seen as an extension of the XY1 phase.
Nevertheless, $K_a$ is renormalized by disorder and $K_o$
flows to infinity. Thus, the function $g$ is equal to zero, 
whereas $h$
is either finite or singular according to the value of $K_a$.
The stability discussion is exactly the same as that of the XY1 
phase apart
from the random XY field. The novelty is that $K_a$ may be
less than 1 but the a-sector is still massless so  there is a
transition line in the 
${\mathcal D}_{PF}^{(0)}-K_a$ plane,
corresponding to a critical value of $(K_a )_c$=1/12. For 
$K_a > (K_a )_c$,
the line we consider is unstable under weak PF disorder. 
The line starts from the XY1 boundary for which
$K_a = 1$ and crosses the frontier line between AF
and H phases. The stiffness $K_a$ decreases along the line 
but we do not know if it reaches the value 1/12 before arriving at 
the tricritical point. We conclude that the line is unstable under 
a random XY field at least in the neighborhood of the XY phase. 

\subsection{XY2 phase}

This phase is characterized by the following IR values in the pure
case~: $K_a^* >1$ and $K_o^* =0$. 
We are in a most interesting case in which the operator $ZF^{(\pi)}$
may be either relevant or irrelevant within the bulk of this phase
according to the scaling equation (\ref{de1}).
We first consider the influence of the
random z-field~: we just need to consider the
a-sector and the operator $ZF^{(\pi)}$, which
simplifies the calculation and does not change the result 
(the random forward scattering $ZF^{(0)}$ does not change the global
picture). The {\it effective} renormalization group equations are 
then at  lowest order
(since the function $g$ is not singular near the IR pure 
fixed point, nothing new happens at the next order)~: 
\begin{mathletters}
\begin{eqnarray} 
&& 
\frac{dy_1}{dl} = 2(1-K_a) y_1 ,
\\ 
&&
\frac{d{\mathcal D}_{ZF}^{(\pi)}}{dl} =
 (3- K_a ){\mathcal D}_{ZF}^{(\pi)} ,
\\ 
&& 
\frac{dK_a}{dl} = -( {\mathcal D}_{ZF}^{(\pi)} 
+ \frac{2}{\pi^2} y_1^2) K_a^2 .
\end{eqnarray} 
\end{mathletters}
In each plane $y_{1}=0$ and ${\mathcal D}_{ZF}^{(\pi)} =0$, 
the flow has a simple form. 
We can draw the corresponding renormalization flow
in the three dimensional space~: it is given in Figure~3.
We observe that the $ZF^{(\pi)}$ operator breaks the XY2 phase 
into two phases. The first one (for large K) is a massless phase 
stable under
small disorder, and the interaction term also flows to zero; the
other phase is unstable under weak disorder. Thus the
XY2 phase is only {\it partially} 
stable under small disorder, and this is
consistent with physical intuition. Indeed, when the single-ion
$D(S^z_i)^2$ term becomes large and negative, we expect that the S=1 
spin chain will behave as the S=1/2 spin chain. As shown by Doty and
Fisher for the S=1/2 spin chain, there is a region of exchange
parameter $\Delta$ stable under weak disorder (it is also a 
superfluid phase arising for the disordered boson gas; see 
Giamarchi and Schulz). In the S=1/2 spin chain, this stability
region is located in the interval $-1<\Delta <-1/2$. 
In the S=1 case, we find that such a stable phase also exists
for K large enough, and this is consistent with the fact that
we should recover the S=1/2 behavior. 

In the case of a random XY field, since there is a random z-field
generated we are in the same situation as above. For the
random XY symmetric coupling, the $PE^{(0)}$
operator is irrelevant and the above discussion is again valid.

In the random XY symmetry-breaking case, the
$PA^{(0)}$ operator is always relevant whereas the $a$-sector is 
ungapped. Thus, the chain is unstable. 
This is again consistent in the limit of large negative $D$
with the Doty and Fisher results for the S=1/2 chain. 


Finally, we briefly comment the case of random z-exchange
that we have not considered yet in this paper.
In terms of the S=1 spins, it is given by~: 
${\mathcal H}_{ZE}=\sum_i J_i S_i^z S_{i+1}^z$. 
When written with the two kinds of spins S=1/2, terms of the form 
$\sigma \tau$ appear. They lead to 
$\cos(\sqrt{8\pi}\phi_a) + \cos(\sqrt{8\pi}\phi_o)$ 
when expressed in the boson language, a coupling that has not been
studied in the context of spin ladders.
The $\cos(\sqrt{8\pi}\phi_a)$ leads to an
operator which is relevant in Haldane phase, 
and make the forward scattering $ZF^{(0)}$ singular and infinite
with the consequence of vanishing interactions in the a-sector.
The AF phase is also unstable~: the $ZF^{(\pi)}$ operator
is also contained in the 
boson expression of this kind of disorder, as if a random z-field 
was generated. These instabilities clearly deserve more studies.
This problem appears also in the S=1/2 
case\cite{DotyFisher}. The XY1 phase is stable while XY2
is unstable.

\section{Conclusion}

We have performed a renormalization group study of the influence of
disorder on the effective theory which describes the S=1 chain. 
The RG equations allow a discussion of
the stability of the various phases of the system. Our discussion
includes the effects of the second order of the renormalization group
calculation. This allows us to capture the Imry-Ma destruction
of the AF phase.

The gapped Haldane and large-D phases are stable under all kinds of
disorder we have studied: random z-field, random XY field and random
XY symmetric and antisymmetric exchange. 
On the contrary, the AF phase which is also gapped is
less stable. In fact, our RG calculation that goes up to the second
order in the disorder strength shows that the phase is unstable to 
an arbitrarily weak random z-field, in agreement with the Imry-Ma 
argument. It is also unstable under random XY-field and
random planar anisotropy since these perturbations do generate
a random z-field. 

Concerning the gapless XY phases, then again the situation is quite
rich. The XY1 phase is stable under random z-field
and random planar exchange while it is
immediately unstable under perturbations breaking the planar symmetry~: 
random XY-field and planar anisotropy. 
The XY2 phase is only partially stable under random fields
and random planar exchange.
In this case the phase breaks
up into two parts: there is a stable massless domain with irrelevant
disorder and another domain with relevant disorder.
Finally XY2 is totally unstable under random planar anisotropy.

With respect to the closely related problem of 2-leg spin ladder,
the effective theory is the same but the the PE and PA type of
disorder leads to different operators
in the bosonization approach. As a consequence,
there are some instabilities that are induced by a different 
mechanism~: this happens in the massless phases XY1 and XY2 
and also along the line separating the Haldane and large-D phases.

Of course it remains to characterize more completely the phases in
which the disorder is relevant. In our calculation, we can simply
observe that the system flows to strong coupling but the methods we
use are not informative on its fate. An outstanding problem is to
describe in a unified manner the weak-coupling regime we observe here
with the regime probed by the real-space RG calculations. The gapless
phase with hidden long-range order of the random spin-1
chain\cite{HY,MGJ} appears presumably only beyond some critical
strength and it seems to be out of reach of the methods we have used
in this paper.

{\it Note added:} A recent work by Y. Nishiyama on the effect of a 
random z-field on the Haldane gap and the XY1 phase is in agreement with our findings (see e-print cond-mat/9805110)

\acknowledgments

We thank M. Bauer, D. Bernard, O. Golinelli, C. Monthus for useful
discussions. We thank also S. Fujimoto for discussions about
the S=1/2 case.

\appendix
\section{stability of a gapped phase}

The pure SG system is gapped when $K \longrightarrow 0$. We call
$K^0$ and $y^0$ a solution of the pure differential system. We set
$K=K^0 + \delta K$ and $y=y^0 + \delta y$. A straightforward
calculation leads to~: 
\begin{eqnarray}
 && 
 \delta K^{\prime} = D-2y^0
(K^0)^2 \delta y - 2(y^0)^2 K^0 \delta K \\
 && 
 \delta y^{\prime} =
(2-a^2K^0)\delta y - a^2 y^0 \delta K 
\end{eqnarray}
From the flow equations, we have
$D(l) < D_0 e^{3l}$, $\delta K > 0$ and $\delta y >0$. Furthermore, 
the
initial values of the functions $\delta K$ and $\delta y$ are 0. 
From this statement, we can solve exactly the
differential system for $\delta K$ and $\delta y$ if we replace 
$D(l)$ by
 $D_0 e^{3l}$. We get~: 
\begin{eqnarray} 
&& 
\delta K(l) =
D_0 F(l) \\ 
&& 
\delta y(l) = D_0 G(l) 
\end{eqnarray} 
where the
functions F and G obey the following differential system~:
\begin{eqnarray} 
&& 
F^{\prime}(x) = e^{3x} - 2y^0(x)^2 K^0(x) F(x) -
2y^0 (x) K^0(x)^2 \\
 && 
 G^{\prime}(x) = (2-a^2 K^0(x)) G(x) -a^2
y^0(x) F(x) \\ 
&& 
F(0) = G(0) = 0. 
\end{eqnarray}
If we choose a small $\epsilon > 0$, it is possible to
find $l=l_0$ so that $K^0(l_0)<\epsilon / 2$. Moreover, we know
that the real function $\delta K$ is inferior to $D_0 F(x)$ where
$F(x)$ is a known function; so we can choose the initial value of D(l)
so that $\delta K(l_0) < \epsilon / 2 $. In this case we have $K(l_0)
< \epsilon$. Because of the positivity of the functions $\delta K$
and $\delta y$, we know that $y^2(l) K^2(l) > (y^0)^2(l) (K^0)^2(l)$
for all l, and so $dK / dl < - (y^0)^2(l) (K^0)^2(l) + D_0 e^{3l}$.
This inequality proves that for sufficiently small $D_0$, we can also
have $K^{\prime}(l_0) < 0$. This argument shows that, for all
$\epsilon > 0$, there is a positive  $l_0$ and a small
enough $D_0$ so that $K(l_0) < \epsilon$ and $K^{\prime}(l_0) < 0$.
 For $l=l_0$, $3-b^2 /K(l_0) \sim -b^2 /K(l_0)$. As a result, the
disorder contribution in $K^{\prime}$ will decrease to 0 as fast as
$e^{-b^2/K(l)l}$, whereas the other term will decrease slower than
$K^2(l)$. Thus this last term dominates the value of $K^{\prime}$
which remains negative for all $l>l_0$. It follows that for $l \gg
l_0$, we have $K^{\prime} \sim -y^2 K^2$, so that $K(l)\rightarrow 0$
for a sufficiently small but finite value of $D_0$.



\begin{figure} 
\epsfysize=10cm $$ \epsfbox{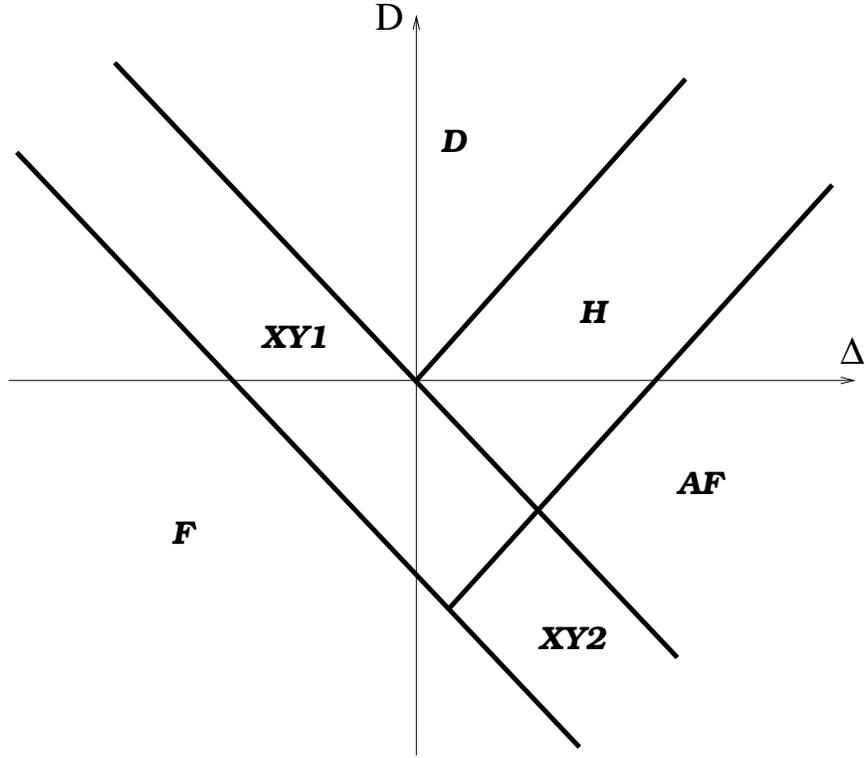} $$
\caption{Spin-1 chain phase diagram without randomness} 
\end{figure}

\begin{figure} 
\epsfxsize=10cm $$ \epsfbox{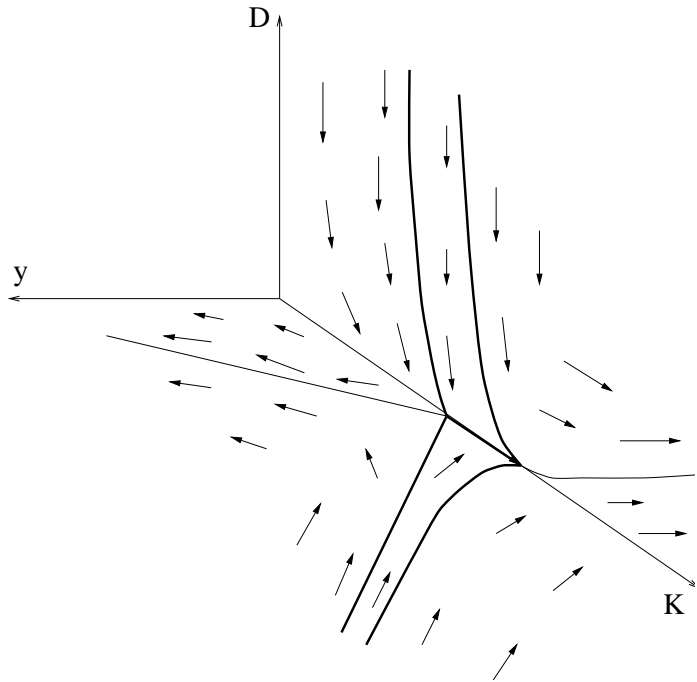} $$
\caption{renormalization flow in the case $(ab)^2 < 6$} 
\end{figure}

\begin{figure} 
\epsfxsize=10cm $$ \epsfbox{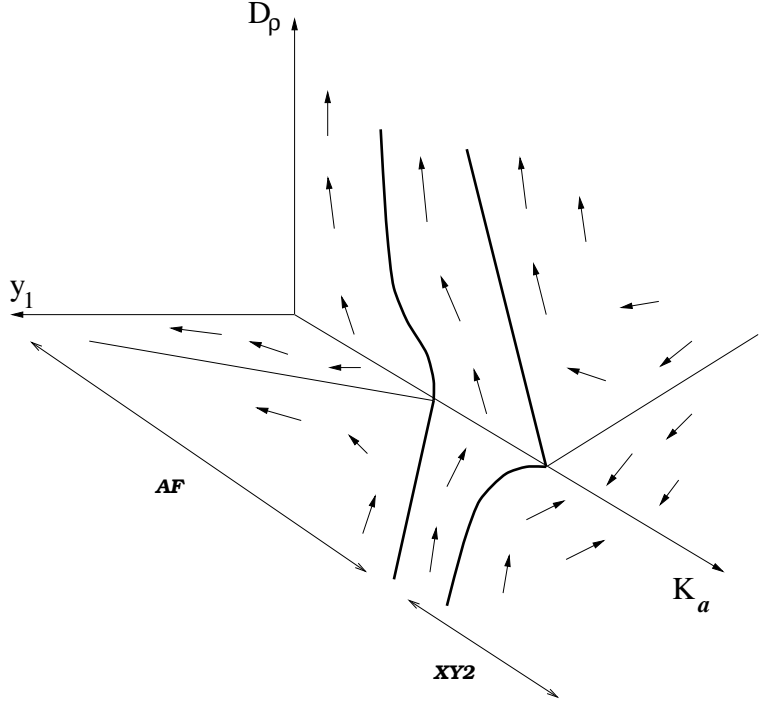} $$
\caption{Effective renormalization flow for AF and XY2 phases}
\end{figure}

\vfill
\eject


\begin{table} 
\begin{tabular}{|c||*5{c|}}
&AF&Haldane&large-D&XY1&XY2\\ 
\hline
 \hline 
 $<\phi_a>$&0&0&$\pi/\sqrt{8}$&dis&dis
\\ 
\hline
 $<\phi_o>$&0&dis&dis&dis&0
 \\ 
\hline
$<\tilde{\phi}_o>$&dis&0&0&0&dis
\\ 
\hline 
$K_a^*$&0&0&0&$>1$&$>1$
\\
\hline 
$K_o^*$&0&$\infty$&$\infty$&$\infty$&0
\\ 
\end{tabular}
\caption{Values of the different parameters characterizing the phases
of the pure S=1 antiferromagnetic chain.} 
\end{table}

\begin{table}
\begin{center} 
\begin{tabular}{|c|c|c|}
 &$q=0$&$q=\pi$
 \\  
 \hline 
 \hline
 ZF&
 $ \partial_x 
\phi_a$
&
$\cos(\sqrt{2\pi}\phi_a) \, \cos ( \sqrt{2 \pi} \phi_o) $
\\ 
\hline
 PF&
  $\cos (\sqrt{\pi / 2} \tilde{\phi_a}) \, 
 \cos(\sqrt{\pi / 2}\tilde{\phi_o})$
 &
 $\cos ( \sqrt{2\pi}\phi_a ) \, \cos (\sqrt{2 \pi} \phi_o)$
 \\ 
 \hline
 PE&
 $\cos ( \sqrt{2\pi}\tilde{\phi_o})$&
 $\cos ( \sqrt{2\pi}\phi_a ) \, \cos (\sqrt{2 \pi} \phi_o)$
  \\ 
  \hline
  PA& 
  $\cos(\sqrt{2\pi}\tilde{\phi_a})$&
  $\cos ( \sqrt{2\pi}\phi_a ) \, \cos (\sqrt{2 \pi} \phi_o)$
   \\ 
\end{tabular}
\caption{Bosonized formulas of the quenched disorder.} 
\end{center}
\end{table}

\begin{table} 
\begin{center}
\begin{tabular}{|c|c|c|c|c|c|}
 &AF&Haldane&large-D&XY1&XY2
 \\ 
\hline 
\hline
ZF&Unstable&Stable&Stable&Stable&Partially stable
\\
\hline
PF&Unstable&Stable&Stable&Unstable&Partially stable
\\
\hline
PE&Unstable&Stable&Stable&Stable&Partially stable
\\
\hline
PA&Unstable&Stable&Stable&Unstable&Unstable
\\ 
\end{tabular}
\caption{Stability of the S=1 phases with respect 
to quenched disorder.} 
\end{center}
\end{table}


\begin{references}

\bibitem{FDM} 
F. D. M. Haldane, 
Phys. lett. {\bf 93A}, 464 (1983);
Phys. Rev. Lett. {\bf 50}, 1153 (1983).

\bibitem{schulz} 
H. J. Schulz, 
Phys. Rev. B{\bf 34}, 6372 (1986).

\bibitem{DNR} 
M. den Nijs and K. Rommelse, 
Phys. Rev. B{\bf 40}, 4709 (1989).

\bibitem{SM} 
S. Strong and A. J. Millis, 
Phys. Rev. Lett. {\bf 69}, 2419 (1992); 
Phys. Rev. B{\bf 50}, 9911 (1994).

\bibitem{STN} 
D. G. Shelton, A. M. Tsvelik and A. A. Nersesyan, 
Phys. Rev. B{\bf 53}, 8521 (1996).

\bibitem{Fisher} 
D. S. Fisher, 
Phys. Rev. B{\bf 50}, 3799 (1994);
Phys. Rev. B{\bf 51}, 6411 (1995).

\bibitem{MDH} 
S. K. Ma, C. Dasgupta, and C. K. Hu, 
Phys. Rev. Lett. {\bf 43}, 1434 (1979); 
C. Dasgupta and S. K. Ma, 
Phys. Rev. B{\bf 22}, 1305 (1980).

\bibitem{BhattLee} 
R. N. Bhatt and P. A. Lee, 
J. Appl. Phys. {\bf 52}, 1703 (1981); 
Phys. Rev. Lett. {\bf 48}, 344 (1982).

\bibitem{DotyFisher} 
C. A. Doty and D. S. Fisher, 
Phys. Rev. B{\bf 45}, 2167 (1992).

\bibitem{West} 
E. Westerberg, A. Furusaki, M. Sigrist, and P. A. Lee,
Phys. Rev. Lett. {\bf 75}, 4302 (1995); 
preprint cond-mat/9610156.

\bibitem{Ha} 
R. A. Hyman, K. Yang, R. N. Bhatt, and S. M. Girvin,
Phys. Rev. Lett. {\bf 76}, 839 (1996).

\bibitem{Hb} 
K. Yang, R. A. Hyman, R. N. Bhatt, and S. M. Girvin, 
J. Appl. Phys. {\bf 79}, 5096 (1996).

\bibitem{Hida} 
K. Hida, e-print cond-mat/9612232.

\bibitem{AKLT} 
I. Affleck, T. Kennedy, E. H. Lieb and H. Tasaki,
Phys. Rev. Lett. {\bf 59}, 799 (1987).

\bibitem{Boe} 
B. Boechat, A. Saguia and M. A. Continentino, 
Solid State Commun. {\bf 98}, 411 (1996).

\bibitem{HY} 
R. A. Hyman and K. Yang, 
Phys. Rev. Lett. {\bf 78}, 1783 (1997).

\bibitem{MGJ} 
C. Monthus, O. Golinelli and Th. Jolic\oe ur, 
Phys. Rev. lett. {\bf 79}, 3254 (1997);
e-print cond-mat/97, to appear in Phys. Rev. B.

\bibitem{Giam} 
T. Giamarchi and H. J. Schulz, 
Phys. Rev. B{\bf 37}, 325 (1988); 
J. Phys. (Paris) {\bf 49}, 819 (1988).

\bibitem{Fuji} 
S. Fujimoto and N. Kawakami, 
Phys.  Rev.  {\bf B 54}, 11018 (1996). 

\bibitem{ImryMa} 
Y. Imry and S. K. Ma,
Phys. Rev. lett. {\bf 35}, 1399 (1975).

\bibitem{Ori}
E. Orignac and T. Giamarchi, 
Phys. Rev. B{\bf 57}, 5812 (1998).

\bibitem{Luther}
J. Timonen and A. Luther,
J. Phys. C{\bf 18}, 1439 (1985).

\bibitem{Boso} 
V. J. Emery, in {\it Highly Conducting One-Dimensional
Solids}, ed. by J. T. DeVreese, R. E. Evrard and V. E. Van Doren,
Plenum Press, New York (1979). See also I. Affleck, in {\it Fields,
Strings and Critical Phenomena}, ed. by E. Br\'ezin and J.
Zinn-Justin, Elsevier Science Publishers B. V., Amsterdam (1989).

\bibitem{dN} 
M. P. M. den Nijs, 
Physica {\bf 111}A, 273 (1982).


\bibitem{Pang} 
H. Pang, S. Liang and J. F. Annett, 
Phys. Rev. Lett. {\bf 71}, 4377 (1993).

\bibitem{Haas} 
S. Haas, J. Riera and E. Dagotto, 
Phys. Rev. B{\bf 48}, 13174 (1993).

\bibitem{Runge} 
K. J. Runge and G. T. Zimanyi, 
Phys. Rev. B{\bf 49}, 15212 (1994).

\bibitem{Eckern} 
P. Schmitteckert, T. Schulze, C. Schuster, P. Schwab and U. Eckern,
``Anderson localization versus delocalization of interacting
fermions in one dimension''
e-print cond-mat/9706107.

\end{references}
\end{document}